\documentclass[10pt]{article}

\usepackage{fancyhdr}
\usepackage{extramarks}
\usepackage{amsmath}
\usepackage{amsthm}
\usepackage{amsfonts}
\usepackage{siunitx}
\usepackage{tikz}
\usepackage[plain]{algorithm}
\usepackage{algpseudocode}
\usepackage{multirow}
\usepackage{booktabs}
\usepackage{graphicx}
\usepackage{subfigure}
\usepackage[colorlinks,linkcolor=black,anchorcolor=black,citecolor=black,urlcolor=blue]{hyperref}
\usepackage{amsmath,bm}
\usepackage{booktabs}
\usepackage{mathtools}
\usepackage{amssymb}
\usepackage{caption}
\usepackage{capt-of}
\usepackage{mciteplus}
\usepackage{cite}
\usepackage{mathrsfs}
\usepackage[title,titletoc,toc]{appendix}
\usepackage{xr}
\usepackage{parskip}
\usetikzlibrary{automata,positioning}
\usepackage{wrapfig}
\renewcommand{\part}[1]{\textbf{\large Part \Alph{partCounter}}\stepcounter{partCounter}\\}

\begin{document}

\title{Mathematics-assisted directed evolution and protein engineering}
\author{Yuchi Qiu$^1$ and Guo-Wei Wei 
Guo-Wei Wei$^{1,2,3}$\footnote{
			Corresponding author.		Email: weig@msu.edu} \\
		$^1$ Department of Mathematics, \\
		Michigan State University, MI 48824, USA.\\
		$^2$ Department of Electrical and Computer Engineering,\\
		Michigan State University, MI 48824, USA. \\
		$^3$ Department of Biochemistry and Molecular Biology,\\
		Michigan State University, MI 48824, USA. \\
	}
\maketitle

\begin{abstract} 
Directed evolution is a molecular biology technique that is transforming protein engineering by creating proteins with desirable properties and functions. However, it is experimentally impossible to perform the deep mutational scanning of the entire protein library due to the enormous mutational space, which scales as $20^N$, where $N$ is the number of amino acids. This has led to the rapid growth of AI-assisted directed evolution (AIDE) or AI-assisted protein engineering (AIPE) as an emerging research field. 
  Aided with advanced natural language processing
(NLP) techniques, including long short-term memory, autoencoder, and transformer, sequence-based embeddings have been   dominant approaches in AIDE and AIPE. Persistent Laplacians, an emerging technique in topological data analysis (TDA), have made structure-based embeddings a superb option in AIDE and AIPE. We argue that a class of persistent topological Laplacians (PTLs), including persistent
Laplacians, persistent path Laplacians, persistent sheaf Laplacians, persistent hypergraph Laplacians,
persistent hyperdigraph Laplacians, and evolutionary de Rham-Hodge theory, can effectively overcome
the limitations of the current TDA and offer a new generation of more powerful TDA approaches. In the general framework of topological deep learning, mathematics-assisted directed evolution (MADE) has a great
potential for   future protein engineering.

\end{abstract}

keywords: Directed evolution, protein engineering, topological data analysis, persistent topological Laplacians, topological deep learning. 

\begin{wrapfigure}{r}[0.0in]{3.2in} 
	\vspace{-3mm}
	\includegraphics[keepaspectratio,width=3.2in]{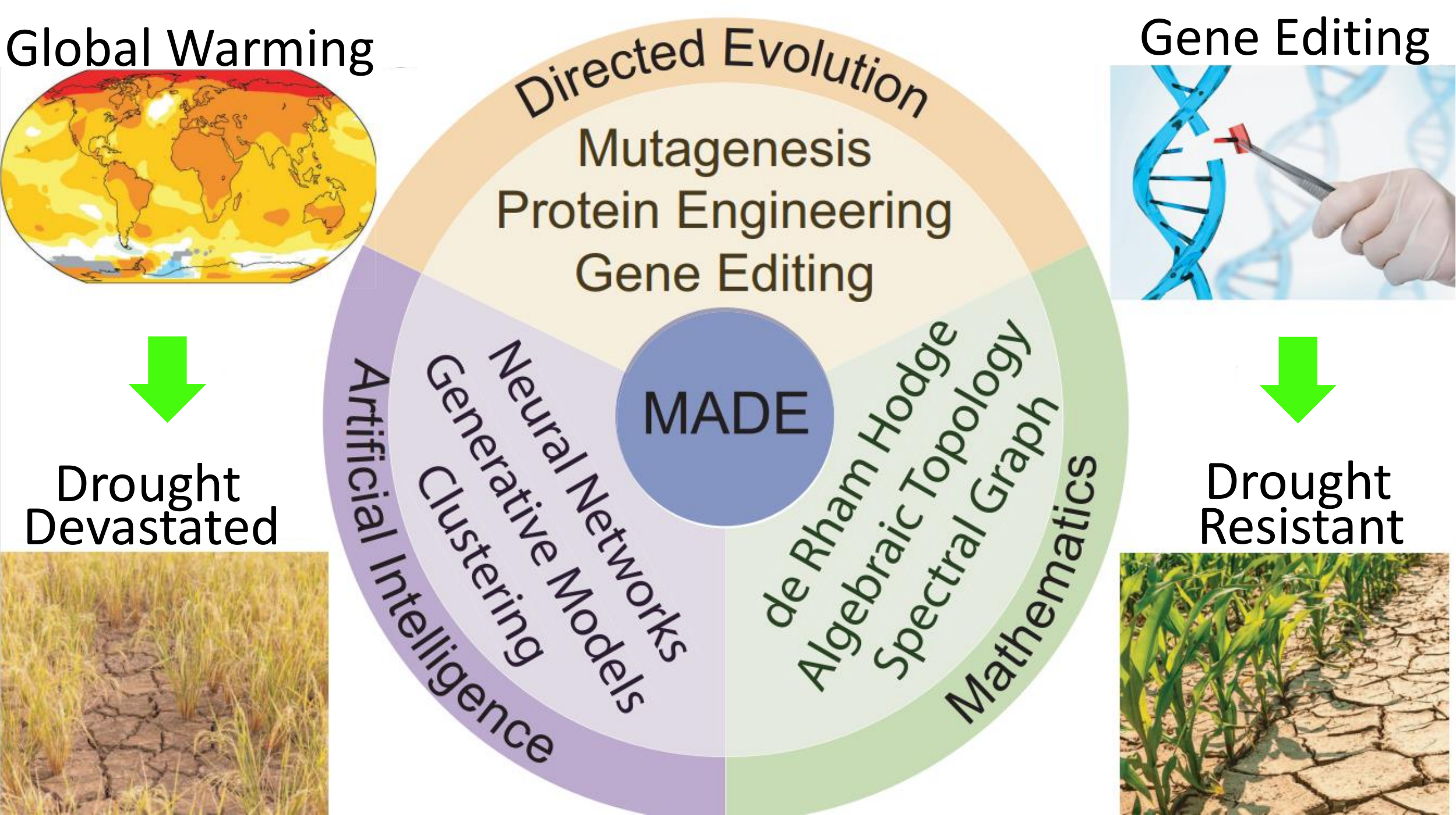}
	\vspace{-5mm}
    \caption{Illustration of mathematics-assisted directed evolution (MADE). Image source: (top left) \url{https://bio.libretexts.org/Courses/Gettysburg_College/01\%3A_Ecology_for_All/24\%3A_Human_Global_Environment/24.02\%3A_Implications_of_Climate_Change}; (top right) \url{https://autism.org/funded-research-2010-2011/}; (bottom left) \url{https://www.shutterstock.com/search/drought-rice}; (bottom right) \url{https://news.mit.edu/2017/climate-change-drought-corn-yields-africa-0316}.
		}
    \label{fig:MADE}
\end{wrapfigure}

\newpage
Climate change is resulting in more frequent and intense heatwaves, droughts, and extreme weather events, rising sea levels, loss of biodiversity, and more severe natural disasters. These changes are having significant impacts on human societies, economies, food security, and the environment. Directed evolution, a powerful molecular biology technique that can engineer organisms for environmental remediation (see Figure \ref{fig:MADE}), may be able to mitigate some of the effects of climate change. Directed evolution has also been applied in protein engineering to design enzymes for industrial processes, develop novel therapeutics, and control harmful pathogens \cite{arnold1998design}. This approach has revolutionized the field of protein engineering, allowing researchers to create proteins with tailored properties and functions that would be difficult or impossible to achieve using traditional methods. However, directed evolution faces significant challenges due to the astronomically large mutational space involved, such as 20$^{200}$ for a relatively small protein with 200 amino acids. Additionally, organisms typically have numerous different proteins, making most directed evolution tasks intractable.

Recently, there has been significant interest in artificial intelligence (AI)-assisted directed evolution (AIDE) in the research community \cite{narayanan2021machine }. AI has proven to be an effective tool for exploring the protein fitness landscape and identifying optimal evolutionary paths \cite{ qiu2021cluster,yang2019machine}. When combined with experimental validation, AIDE offers an alternative to rational protein design, potentially leading to more effective antibodies and enzymes, as well as flood-, drought-, and disease-resistant plants. Furthermore, AIDE has contributed to a deeper understanding of evolutionary principles. Although AIDE primarily concentrates on protein sequence data, structure data offers a wealth of biophysical information that cannot be extracted from sequences. The main challenge in utilizing structure data stems from its structural complexity, high dimensionality, nonlinearity, and the multiscale and multiphysical interactions inherent in biological systems.

In recent years, the impact of topological data analysis (TDA) on science and engineering has grown exponentially \cite{kaczynski2004computational,wasserman2018topological}. The main tool of TDA, persistent homology (PH) \cite{edelsbrunner2008persistent, zomorodian2004computing}, bridges the gap between complex geometry and abstract topology through filtration. Represented in terms of  persistence barcodes \cite{ghrist2008barcodes}, persistence images \cite{adams2017persistence}, and/or persistence landscapes \cite{bubenik2015statistical}, PH has been incredibly successful in handling intricately complex, high-dimensional, nonlinear, and multiscale data \cite{cang2017topologynet, nguyen2019mathematical,townsend2020representation}, including those from computer-aided drug design \footnote{\url{https://sinews.siam.org/Details-Page/persistent-homology-analysis-of-biomolecular-data}} and accurate prediction of viral future mutation sites \cite{chen2020mutations}. However, it has limitations, including its inability to handle heterogeneous information (i.e., different types of atoms in proteins), its qualitative nature (e.g., ignoring the difference between a 5-member ring and a 6-member ring), its lack of description of non-topological changes (i.e., homotopic shape evolution), its incapability of coping with directed networks and digraphs (i.e., polarization, gene regulation), and its inability to characterize structured data (e.g., functional groups and protein domains and motifs) \cite{wei2023topological}. 

\begin{wrapfigure}{r}[0.0in]{5.1in} 
	\vspace{-5mm}
	\includegraphics[keepaspectratio,width=5.1in]{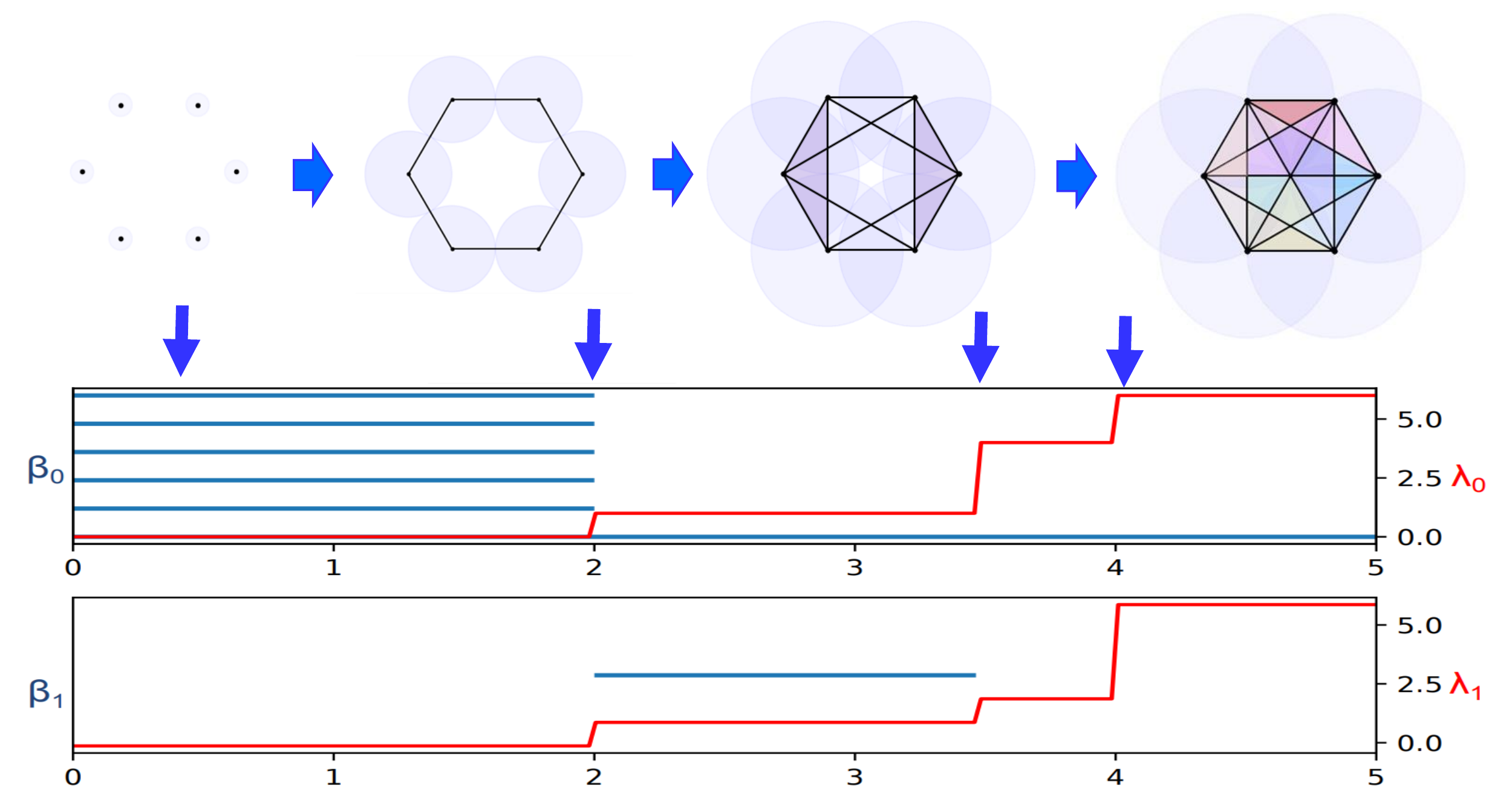}
	\vspace{-5mm}
    \caption{Comparison of persistent homology (PH) \cite{edelsbrunner2008persistent, zomorodian2004computing} and persistent Laplacians (PLs) \cite{wang2020persistent} for  six points. 
The filtration  characterized by the horizontal axis $r$ of six points is shown in the top chart. 		
The corresponding topological features of dimension 0 and dimension 1 are shown the second and third charts, respectively. 
PH barcodes ($\beta_0(r)$ and $\beta_1(r)$) are given in blue. 
The first non-zero eigenvalues of dimension 0 ($\lambda_0(r)$) and dimension 1 ($\lambda_1(r)$) of PLs are depicted in red.
The harmonic spectra of PLs return all the topological invariants of PH, whereas the non-harmonic spectra of PLs capture the additional homotopic shape evolution of PLs during the filtration that are neglected by PH. 
		Figure courtesy of Gengzhuo Liu. 
		}
    \label{fig:Laplacians}
\end{wrapfigure}

To address these challenges, many persistent topological Laplacians (PTLs) have been introduced, including persistent Laplacians \cite{wang2020persistent, memoli2022persistent},
persistent path Laplacians \cite{wang2023persistent}, 
persistent sheaf Laplacians \cite{wei2021persistent}, 
persistent hypergraph Laplacians \cite{liu2021persistent}, persistent hyperdigraph Laplacians \cite{chen2023persistent}, and evolutionary de Rham-Hodge theory \cite{chen2021evolutionary}. Among them, persistent Laplacians, also known as persistent spectral graphs, can capture the homotopic shape evolution of data that cannot be described by PH (see Figure \ref{fig:Laplacians}).  Topological Laplacians are originated from the de Rham-Hodge theory in differential geometry.  Eckmann  introduced simplicial complex to classical graph Laplacians in 1944 \cite{eckmann1944harmonische}, resulting in topological Laplacians on graph, which enable the description of many-body interactions. The introduction of  persistence  by Wang et al \cite{wang2020persistent} endows topological Laplacians an excellent tool for topological deep learning, an emerging paradigm in data science \cite{pun2022persistent}.   Persistent Laplacians instrumented the accurate forecasting of emerging dominant SARS-CoV-2 variants BA.4/AB.5 \cite{chen2023persistent}   \footnote{\url{https://sinews.siam.org/Details-Page/topological-artificial-intelligence-forecasting-of-future-dominant-viral-variants}}. Persistent path Laplacians were designed for directed graphs and directed networks \cite{wang2023persistent}. They can be regarded as a generalization of path complex \cite{grigor2020path} and persistent path homology \cite{chowdhury2018persistent}.
 Persistent sheaf Laplacians, motivated by sheaf Laplacians \cite{hansen2019toward},  enable the embedding of physical laws into topological invariants and the description of heterogeneous properties \cite{wei2021persistent}. 
Persistent hypergraph Laplacians allow a topological description of internal organizations in the data \cite{liu2021persistent}. 
 Persistent hyperdigraph Laplacians can further deal with directed topological hypergraphs \cite{chen2023persistent}. Finally, defined on a family of filtration-induced differentiable manifolds,  evolutionary de Rham-Hodge theory or persistent Hodge Laplacians,  uniquely provide a multiscale topological analysis of volumetric data \cite{chen2021evolutionary}. Persistent Hodge Laplacians can be regarded as the continuum counterpart of  (discrete) persistent Laplacians due to their similarity in the involved algebraic topology structures. However, their underlying mathematical definitions, i.e., differential forms on manifolds and simplicial complexes on graphs, are completely different. Note that these topological Laplacians differ further from graph Laplacians, which describe pairwise interactions among nodes and do not have high-dimensional topological structures as topological Laplacians do.      
 Alternatively, quantum persistent homology or persistent Dirac also offers a solution to some of the current limitations of PH \cite{ameneyro2022quantum, wee2023persistent}.

The effectiveness and practicality of new  TDA  tools have been demonstrated in an emerging  topological deep learning paradigm for AIDE and AI-assisted protein engineering (AIPE) research \cite{qiu2023persistent}. Specifically,  Element-specific PH and element-specific persistent Laplacians  were developed to simplify molecular geometric complexes, reduce protein dimensionality, and capture topological invariants, shape evolution, and sequence disparities in the protein fitness landscape. These structure embeddings were combined with sequence embeddings obtained from advanced natural language processing (NLP) tools, such as autoencoders, long short-term memory, and transformers \cite{narayanan2021machine}. These methods were further integrated with cutting-edge machine learning algorithms to establish a new generation of    AIDE and AIPE approaches \cite{qiu2023persistent}.
 
Scientists are enthusiastic and optimistic about  mathematics-assisted directed evolution and protein engineering \cite{luo2023sensing}. 
Mathematicians can also uniquely contribute to this emerging field through the development of
  statistical models  and mathematical frameworks that optimize the efficiency and
accuracy of machine learning algorithms, capture the behavior of AI and machine learning systems, provide
the fundamental insights into the capabilities and limitations of AI systems, and analyze the computational
complexity of machine learning algorithms \cite{mucllari2022orthogonal}. 
The continuous advances in mathematical techniques and innovative AI algorithms will determine the future of directed evolution and protein engineering and address  some of the grand challenges of our time, such as climate change, food security, and pandemic prevention.


{\it Yuchi Qiu is a research associate at Michigan State University. His research focuses on AI- and mathematics-assisted directed evolution and protein engineering. 
Guo-Wei Wei is an MSU Foundation Professor at Michigan State University. His research concerns the mathematical foundations of bioscience and data science.}

\end{document}